\documentclass[proof]{WileyASNA-v1}
\articletype{Original Article}%

\received{\today}
\revised{$ $Revision: 1.51 $ $}
\accepted{}

 \usepackage{natbib}
\usepackage{bm}
\graphicspath{{./fig/}{./png/}}

\newcommand{\EQ}{\begin{equation}}
\newcommand{\EN}{\end{equation}}
\newcommand{\EQA}{\begin{eqnarray}}
\newcommand{\ENA}{\end{eqnarray}}

\newcommand{\Eq}[1]{Equation~(\ref{#1})}

\newcommand{\EEqss}[2]{Equations~(\ref{#1})--(\ref{#2})}

\newcommand{\Fig}[1]{Figure~\ref{#1}}

\newcommand{\bra}[1]{\langle #1\rangle}

{}
{}
{}
{}
{}

{}
{}
{}
{}
{}
{}
{}
{}
{}
{}
{}
{}
{}
{}
{}
{}
{}
{}
{}
{}
{}

{}
{}
{}

{}

{}
{}

\newcommand{\kk}{\bm{k}}

\newcommand{\xx}{\bm{x}}

\newcommand{\BB}{\bm{B}}

\newcommand{\uu}{\mbox{\boldmath $u$} {}}
\newcommand{\UU}{\mbox{\boldmath $U$} {}}

\newcommand{\JJ}{\mbox{\boldmath $J$} {}}

\newcommand{\AAA}{\mbox{\boldmath $A$} {}}

\newcommand{\ee}{\mbox{\boldmath $e$} {}}

\newcommand{\ff}{\mbox{\boldmath $f$} {}}

\newcommand{\nab}{\mbox{\boldmath $\nabla$} {}}

\newcommand{\SSSS}{\mbox{\boldmath ${\sf S}$} {}}

\newcommand{\ii}{{\rm i}}

\newcommand{\DDn}{{\rm D}}
\newcommand{\DD}{{\rm D} {}}

\newcommand{\dd}{{\rm d} {}}

\newcommand{\const}{{\rm const}  {}}

\def\Pm{\mbox{\rm Pr}_M}
\def\Rm{\mbox{\rm Re}_M}

\def\Lu{\mbox{\rm Lu}}

\def\EEK{{\cal E}_{\rm K}}
\def\EEM{{\cal E}_{\rm M}}
\def\EK{E_{\rm K}}
\def\EM{E_{\rm M}}

\def\cs{c_{\rm s}}

\def\vA{v_{\rm A}}
\def\vAz{v_{\rm A0}}

\def\kf{k_{\rm f}}

\def\EM{E_{\rm M}}

\def\epsM{\epsilon_{\it M}}

\def\urms{u_{\rm rms}}

\def\brms{b_{\rm rms}}

\def\half{{\textstyle{1\over2}}}

\def\onethird{{\textstyle{1\over3}}}

\newcommand{\s}{\,{\rm s}}

\newcommand{\yapj}[3]{: #1, {\it ApJ} {\it #2}, #3}
\newcommand{\yapjl}[3]{: #1, {\it ApJ} {\it #2}, #3}

\newcommand{\yrpp}[3]{: #1, {\it RPPh} {\it #2}, #3}
\newcommand{\yana}[3]{: #1, {\it A\&A} {\it #2}, #3}
\newcommand{\yanar}[3]{: #1, {\it A\&AR} {\it #2}, #3}

\newcommand{\ygafd}[3]{: #1, {\it GApFD} {\it #2}, #3}
\newcommand{\yjfm}[3]{: #1, {\it JFM} {\it #2}, #3}
\newcommand{\ynjp}[3]{: #1, {\it NJPh} {\it #2}, #3}

\newcommand{\ypfb}[3]{: #1, {\it PhFlB} {\it #2}, #3}

\newcommand{\yphy}[3]{: #1, {\it Physica} {\it #2}, #3}

\newcommand{\yprl}[3]{: #1, {\it PhRvL} {\it #2}, #3}
\newcommand{\ypra}[3]{: #1, {\it PhRvA} {\it #2}, #3}
\newcommand{\yprd}[3]{: #1, {\it PhRvD} {\it #2}, #3}
\newcommand{\ypre}[3]{: #1, {\it PhRvE} {\it #2}, #3}

\newcommand{\ymn}[3]{: #1, {\it MNRAS} {\it #2}, #3}

\newcommand{\ysph}[3]{: #1, {\it Solar Phys.} {\it #2}, #3}

\newcommand{\ypnas}[3]{: #1, {\it Proc. Nat. Acad. Sci.} {\it #2}, #3}

\newcommand{\yjcp}[3]{: #1, {\it J. Comput. Phys.} {\it #2}, #3}
\newcommand{\yjour}[4]{: #1, {\it #2} {\it #3}, #4}

\newcommand{\ybook}[3]{: #1, {\it \it #2} (#3)}

  \usepackage{color}

\begin{document}

\title{Cross-helically forced and decaying hydromagnetic turbulence}

\author[1,2,3,4]{Axel Brandenburg}

\author[5]{Sean Oughton}

\authormark{Brandenburg and Oughton}

\address[1]{\orgdiv{Nordita},
\orgname{KTH Royal Institute of Technology and Stockholm University},
\orgaddress{\state{Stockholm}, \country{Sweden}}}

\address[2]{\orgdiv{Department of Astronomy}, \orgname{Stockholm University},
\orgaddress{\state{Stockholm}, \country{Sweden}}}

\address[3]{\orgdiv{JILA and Laboratory for Atmospheric and Space Physics},
\orgname{University of Colorado},
\orgaddress{\state{Boulder}, \country{USA}}}

\address[4]{\orgdiv{McWilliams Center for Cosmology and Department of Physics},
\orgname{Carnegie Mellon University},
\orgaddress{\state{Pittsburgh}, \country{USA}}}

\address[5]{\orgdiv{Department of Mathematics and Statistics},
\orgname{University of Waikato}, \orgaddress{\state{Hamilton 3240},
\country{NZ}}}

\corres{A. Brandenburg, Nordita,
KTH Royal Institute of Technology and Stockholm University,
10691 Stockholm, Sweden. \email{brandenb@nordita.org}}

\abstract{
We study the evolution of kinetic and magnetic energy spectra in
magnetohydrodynamic flows in
the presence of strong cross helicity.
For forced turbulence, we find weak inverse transfer of kinetic
energy toward the smallest wavenumber.
This is plausibly explained by the finiteness of scale separation
between the injection wavenumber and the smallest wavenumber of
the domain, which here is a factor of 15.
In the decaying case, there is a slight increase at the smallest
wavenumber, which is probably explained by the dominance of kinetic
energy over magnetic energy at the smallest wavenumbers.
Within a range of wavenumbers covering almost an order of magnitude the
decay is purely exponential, which is argued to be a consequence of a
suppression of nonlinearity due to the presence of strong cross helicity.
}


\fundingInfo{University of Colorado.
NSF Astronomy a Astrophysics Grants Program, 1615100.}

\maketitle

        \section{Introduction}

Conservation laws fundamentally affect the cascade properties of
hydrodynamic and magnetohydrodynamic (MHD) turbulence.
This can be seen both in forced and decaying turbulence, but the
effects are often more dramatic in the decaying case.
Decaying MHD turbulence is strongly affected by the
presence of non-vanishing magnetic helicity \citep[see, e.g.,][]{TKBK12,KTBN13}.
The study of decaying MHD turbulence has recently received increased attention
in connection with the study of the decay of primordial magnetic fields during
the early Universe, because the possibility of an inverse cascade
leads to progressively larger length scales of the turbulent magnetic field
and can reach kiloparsec scales at the present time \citep{BEO96},
if it was of sub-horizon scales at the time of the electroweak phase
transition some $10^{-11}\s$ after the big bang; see \cite{DN13} and
\cite{Sub16} for recent reviews.
The possibility of an inverse cascade is believed to be related
to the conservation of magnetic helicity \citep{FPLM75}.
However, since the work of \cite{Wol58}, we know that there is another
important conserved quantity, the cross helicity.
The question arises whether finite cross helicity can have similar effects.
In particular we wish to know whether cross helicity causes a slow-down
of the decay of MHD turbulence and whether there is accelerated growth
of the correlation length when cross helicity becomes important,
as was found by \cite{CHB01}, \cite{BJ04}, and \cite{TKBK12}, for example.
In those cases, it was found that the fractional magnetic helicity,
i.e., the magnetic helicity normalized by the magnetic energy and the
correlation length, grows proportional to $t^{1/2}$, because magnetic
helicity stays constant, the energy decreases like $t^{-1}$, and the
correlation length increases like $t^{1/2}$.
A qualitatively similar behavior has been found in the shell model work
of \cite{FS10}, although in their case the growth of fractional helicity
may have been a consequence of a true instability, whose origin is not
yet understood, however.

As is well known, in homogeneous incompressible MHD turbulence the
normalized cross helicity tends to grow in magnitude, which is
associated with weakening of the nonlinearities in the MHD equations
    \citep[e.g.,][]{DobrowolnyEA80-prl,DobrowolnyEA80-aa,
        GrappinEA82,GrappinEA83,
        MattEA83,PouquetEA86,
        MattEA08-align,ServidioEA08-depress}.
The related question of a slow-down of the growth of cross helicity
has already been partly addressed by \cite{SB09}, who showed that
for the non-helical Archontis flow \citep{Arc00,ADN03}, which is driven
by a forcing that is proportional to $(\sin kz,\sin kx, \sin ky)$,
the mean magnetic field grows at first exponentially and then develops
a slow saturation behavior.
However, this slow-down does not appear to depend on the microphysical
values of kinematic viscosity or magnetic diffusivity.

For completeness, we note that cross helicity can also be
produced in strongly stratified turbulence in the presence of magnetic
fields parallel to the direction of gravity \citep{RKB11}, which also
leads to the growth of magnetic fields at large length scales, which is
suggestive of inverse transfer behavior \citep{BGJKR14}.
In that case, however, cross helicity develops more rapidly and
it also disappears quickly, if the external magnetic field is removed.
This case is inhomogeneous owing to the presence of gravitational
stratification and will not be considered in the present paper.

        \section{Simulation setup}

Our main objective is to compute the decay of MHD turbulence in the
presence of cross helicity.
As initial conditions we take the result of an earlier simulation where
both velocity $\uu$ and magnetic field $\BB$ were driven by the same
forcing function, which ensures that finite cross helicity $\uu\cdot\BB$
is injected into the system.
We chose the injection wavenumber to be large enough so that there
is a chance to see inverse transfer from the forcing wavenumber $\kf$
to the smallest wavenumber of the domain, $k_1$.
The ratio $\kf/k_1$ is referred to as the scale separation ratio, and
we choose $\kf/k_1=15$ in all our simulations.

Our setup for the driven case is identical to that of \cite{BR13},
except that they included also rotation and used a smaller scale separation
ratio of $\kf/k_1=5$.
As in earlier work \citep{KTBN13,TKBK12},
we use an isothermal equation of state so that the pressure $p$ and
the mass density $\rho$ are proportional to each other, i.e., $p=\rho\cs^2$,
where $\cs=\const$ is the isothermal sound speed.
The magnetic field $\BB$, the fluid velocity $\UU$,
and the mass density $\rho$ obey
\EQ
  \frac{\partial\AAA}{\partial t}
       =
    \UU\times\BB
   - \eta\mu_0\JJ
   + \ff_{\rm M},
\label{eq11}
\EN
\EQ
  \frac{\DD\UU}{\DD t}
       =
   - \cs^2\nab\ln\rho
   + \frac{1}{\rho} \JJ \times \BB
   + \frac{1}{\rho} \nab {\bm\cdot} (2\rho\nu\SSSS)
   + \ff_{\rm K},
\label{eq13}
\EN
\EQ
  \frac{\DD \ln\rho} {\DD t}
       =
     - \nab {\bm\cdot} \UU .
\label{eq15}
\EN
Here, $\AAA$ is the magnetic vector potential, so
        $ \nab \times \AAA = \BB$
is the magnetic field, and
        $ \JJ = \nab \times \BB/\mu_0$
is the current density,
where $\mu_0$ is the vacuum permeability,
 $\eta$ is the magnetic diffusivity,
        $ \DDn / \DD t = \partial /\partial t  + \UU {\bm\cdot} \nab$
is the advective time derivative,
        $ {\sf S}_{ij} = \half(U_{i,j}+U_{j,i}) - \onethird\delta_{ij}\nab {\bm\cdot} \UU$
are the components of the trace-less rate of strain tensor,
$\nu$ the kinematic viscosity, and $\ff_{\rm M}$ and $\ff_{\rm K}$
define the magnetic and kinetic forcings, respectively.
These will be specified below.
We consider small Mach numbers, so compressibility effects are negligible.
No DC magnetic field is imposed.

\EEqss{eq11}{eq15} are solved numerically in a cubic domain
of side length $L$ using periodic boundary conditions.
Thus, $k_1 = 2 \pi / L$ is the smallest possible (non-zero) wavenumber.
During the first part, before studying the decay, we drive the system
in a cross-helical fashion such that
        $ \ff_{\rm K} = \nab \times \ff_{\rm M} $
(ignoring a correction factor for units),
with random functions that are $\delta$-correlated in time.

We approximate a forcing that is $\delta$-correlated in time by
adding after each time step of size $\delta t$ the contributions
$\delta t\ff_{\rm M}$ and $\delta t\ff_{\rm K}$ to $\AAA$ and $\UU$,
respectively, and change $\ff_{\rm M}$ and $\ff_{\rm K}$ randomly
from one step to the next \citep{B01}.
Thus, we put
\EQA
 \ff_{\rm M}\!\!&=&\!\!N_{\rm M} \mbox{Re} \{\ff_{\kk(t)} \exp[ \ii \kk(t) {\bm\cdot} \xx + \ii \phi(t)] \},\\
 \ff_{\rm K}\!\!&=&\!\!N_{\rm K} \mbox{Re} \{\ii \kk(t) \times \ff_{\kk(t)} \exp[ \ii \kk(t) {\bm\cdot} \xx + \ii \phi(t)] \},\quad
\label{eq17}
\ENA
where $N_{\rm M}$ and $N_{\rm K}$ are given by
\EQA
 N_{\rm M}&=&{\cal N}_{\rm M} \cs \sqrt{\mu_0 \rho_0 \cs/\kf\delta t},\\
 N_{\rm K}&=&{\cal N}_{\rm K} \cs \sqrt{\cs/\kf\delta t},
\label{eq19}
\ENA
and ${\cal N}_{\rm M}$ and ${\cal N}_{\rm K}$ are dimensionless amplitudes,
$\rho_0$ is the initial mass density, considered to be uniform,
$\kf$ the average forcing wavenumber,
$\delta t$ the size of the time step, and $\ff_{\kk}$ is given by
\EQ
  \ff_{\kk(t)}= \frac{\kk(t) \times \ee(t)}
                     {\sqrt{\kk(t)^2 - \left[\kk(t) {\bm\cdot} \ee(t)\right]^2}},
\label{eq23}
\EN
where $\ee(t)$ is a unit vector which is in the same sense random as
$\kk(t)$ but not parallel to it \citep{HBD04}.
The wavevector $\kk$ and the phase $\phi$ are random functions of time,
i.e., $\kk = \kk(t)$ and $\phi = \phi(t)$, such that their values within
a given time step are constant.
Note that $\nab {\bm\cdot} \ff_{\rm M}=\nab {\bm\cdot} \ff_{\rm K}=0$.
The wavevectors $\kk$ are chosen such that their moduli $k = |\kk|$ lie
in a band of width $\delta k$ around a mean forcing wavenumber $\kf$,
that is, $\kf - \delta k/2  \leq  k \leq \kf + \delta k/2$,
and we choose $\delta k = k_1$.
In the limit of small time steps, which we approach in our calculations, the forcing may be considered as $\delta$-correlated.

When a statistically steady state is reached, we set
$N_{\rm M}=N_{\rm K}=0$ and study in that way decaying turbulence.
We describe the statistically stationary state of our simulations using
the magnetic Prandtl number $\Pm$, the magnetic Reynolds number $\Rm$,
and the Lundquist number $\Lu$,
\EQ
  \Pm = \nu   / \eta,     \quad
  \Rm = \urms / \eta \kf, \quad
  \Lu = \vA   / \eta \kf,
\label{eq23b}
\EN
with $\urms$ and $\vA=\brms/\sqrt{\mu_0 \rho_0}$ being defined using averages
over the full computational volume.
While $\Pm$ is an input parameter, $\Rm$ and $\Lu$ are used as diagnostics.
We analyze our results in terms of kinetic and magnetic energy spectra,
$\EK(k,t)$ and $\EM(k,t)$ that are normalized such that
  $ \int \EK(k,t) \, \dd k = \bra{\UU^2/2} \equiv \EEK $
and
  $ \int \EM(k,t) \, \dd k = \bra{\BB^2/2\mu_0} \equiv \EEM $
are the mean kinetic and magnetic energy densities, respectively.

For our numerical simulations we use the
\textsc{Pencil Code}\footnote{\url{https://github.com/pencil-code},
DOI:10.5281/zenodo.2315093},
which is a high-order public domain code
for solving partial differential equations, including the hydromagnetic
equations given above.
It uses sixth order finite differences in space and the third order
2N-RK3 low storage Runge--Kutta time stepping scheme of \cite{Wil80}.

        \section{Results}

        \subsection{Driven turbulence}

\begin{figure}[t!]\begin{center}
\includegraphics[width=\columnwidth]{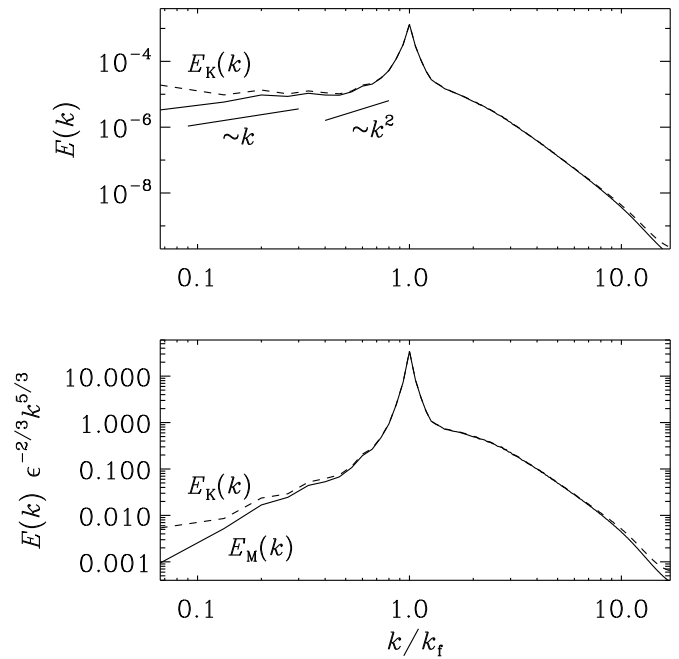}
\end{center}\caption[]{
  Power spectra of kinetic and magnetic energies
  (dashed and dotted lines, respectively) during the driven phase.
  The fractional cross helicity is $\sigma_c \approx 0.81$.
  The lower panel shows spectra compensated by the Kolmogorov scaling
  of $ \epsilon^{-2/3}k^{5/3} $.
  Note the shallow kinetic energy spectrum (dashed line) for small
  values of $k$.
}\label{pspec}\end{figure}

We begin by presenting results of simulations with a resolution of
$ 512^3 $ meshpoints.
In \Fig{pspec} we show kinetic and magnetic energy spectra, in both
uncompensated and compensated forms.
Note that at the smallest wavenumbers ($k/\kf=1/15$) the spectral
kinetic energy exceeds the magnetic spectral energy and its spectrum is
approximately flat.
This is reminiscent of simulations that exhibit an inverse cascade
(or at least inverse transfer) of kinetic energy owing to what is known as the
anisotropic kinetic alpha (AKA) effect of \cite{FSS87}; see also
\cite{BvR01} for simulations at larger Reynolds numbers.
Here, however, no AKA effect is expected to occur.
Thus, the slight uprise of $\EK(k,t)$ at low $k$ is a new result that
occurs now with the addition of cross helicity.
It is therefore tempting to associate it with the conservation of
cross helicity.

        \subsection{Decaying turbulence}

We use a snapshot from the statistically steady phase of
the forced simulation with the forcing being turned off.
During the subsequent decay, the normalized cross helicity,
\begin{equation}
   \sigma_c = \frac {2 \bra{\uu\cdot\BB}} {q\bra{\uu^2}+\bra{\BB^2}/q}
\label{sigmac}
\end{equation}
tracks the evolution of $ \vA / \urms $, both of which
attain peak values reaching 0.996 shortly after turning off the forcing
and then decay to about 0.8 after some 500 sound travel times.
In \Eq{sigmac}, we have introduced $q=\sqrt{\mu_0\rho}$ for
dimensional reasons \citep{ZB18}.

\begin{figure}[t!]\begin{center}
\includegraphics[width=\columnwidth]{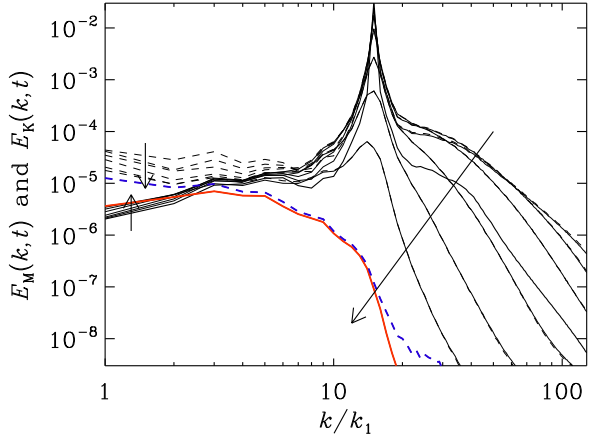}
\end{center}\caption[]{
  Kinetic and magnetic energy spectra (uncompensated)
  at different times during the decay shown as dashed and solid lines, respectively.
  Note that both spectra decrease strongly at
  large wavenumbers. At small wavenumbers the kinetic
  energy decreases in time while the magnetic energy increases,
  thus approaching equipartition.
  The spectra at the last time are shown as a solid red line
  for $\EM(k,t)$ and as a dashed blue line for $\EK(k,t)$.
}\label{pkt}\end{figure}

In \Fig{pkt} we show magnetic and kinetic energy spectra in
logarithmically spaced time intervals during the decay phase
at $t/\cs k_1=5$, 10, 20, 50, 100, 200, and 500 sound travel times,
corresponding to $t/\vAz\kf=1.23$, ..., 123 Alfv\'en times,
where $\vAz$ is the initial Alfv\'en speed.
Note that the spectral kinetic and magnetic energies decrease at large
wavenumbers, while at small wavenumbers the kinetic energy decreases
and the magnetic energy increases so that the two quantities approach
each other at large scales
 (although they are still far from being equal).
We recall that MHD absolute equilibrium studies predict that the
Alfv\'en ratio
        $ r_{\rm A} = \EEK / \EEM \le 1 $
        \citep{StriblingMatt90},
although the relationship of these results to the particular initially
forced situation considered here,
and decaying situations in general
        \citep{StriblingMatt91},
warrants further investigation.
In our case, instead, $r_{\rm A}$ increases from unity to about 1.5
during the course of the simulation.

\begin{figure}[t!]\begin{center}
\includegraphics[width=.99\columnwidth]{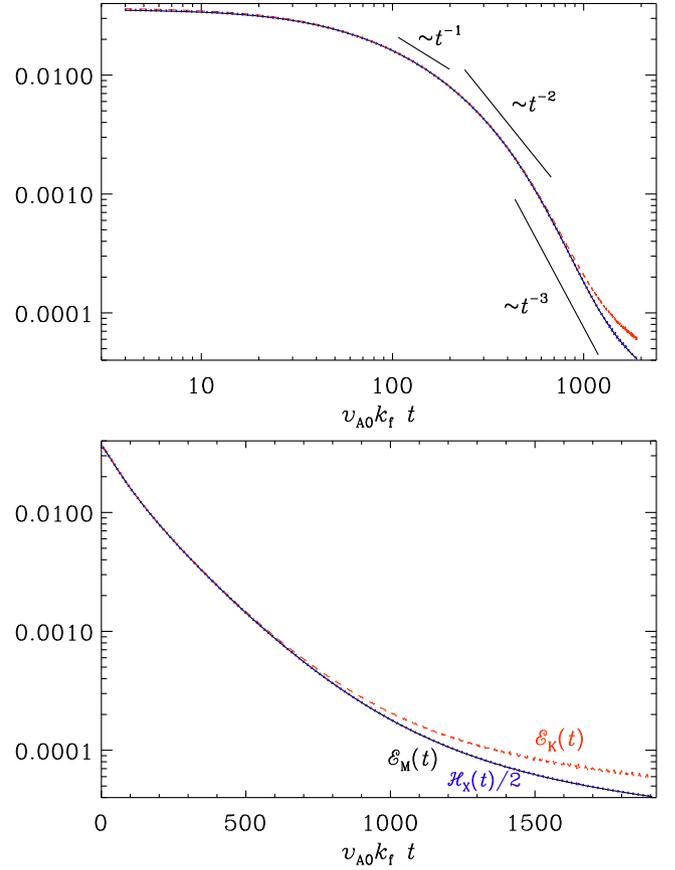}
\end{center}\caption[]{
  Decay of kinetic and magnetic energies.
  The red dashed line indicates kinetic energy and
  the blue dotted line indicates cross helicity.
  The decay follows neither an algebraic (cf.\ upper panel)
  nor an exponential (cf.\ lower panel) decay law.
}\label{pdecay}\end{figure}

The magnetic Prandtl number is unity and the Reynolds number
based on the smallest wavenumber in the domain decreases from
about 5000 to about 200 after about 500 sound travel times.
The decay is neither algebraic nor exponential; see Figure~\ref{pdecay}.
This is explained by the fact that the decay of magnetic and
kinetic energies is actually composed of a continuous sequence
of uncoupled modes, each decaying exponentially with their own resistive
decay rate, $\mu k^2$, where $\mu=\nu+\eta$ is the sum of kinematic
viscosity and magnetic diffusivity.

\begin{figure}[t!]\begin{center}
\includegraphics[width=.9\columnwidth]{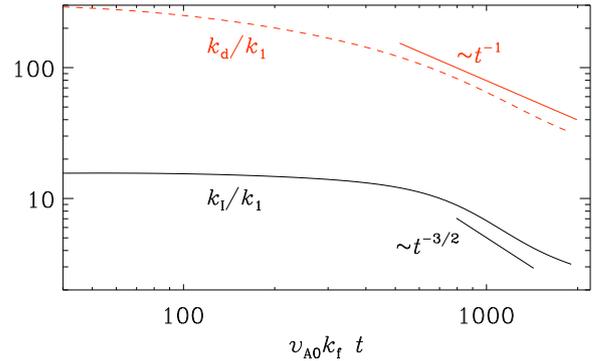}
\end{center}\caption[]{
  Dependence of magnetic integral wavenumber $k_{\rm t}$.
  and the dissipation wavenumber $k_{\rm d}$.
}\label{ppkft}\end{figure}

In \Fig{ppkft} we plot the time dependence of the magnetic
dissipation wavenumber,
\EQ
k_{\rm d}(t)=(\epsM/\eta^3)^{1/4},
\EN
where $\epsM=2\eta\int\EM k^2\,\dd k$ is the instantaneous
mean energy dissipation rate.
We also plot the evolution of the integral wavenumber,
\EQ
k_{\rm I}^{-1}(t)=\left.\int k^{-1}\EM \,\dd k\right/\int \EM \,\dd k.
\EN
Both quantities are evaluated from the actual spectra.

\begin{figure}[t!]\begin{center}
\includegraphics[width=\columnwidth]{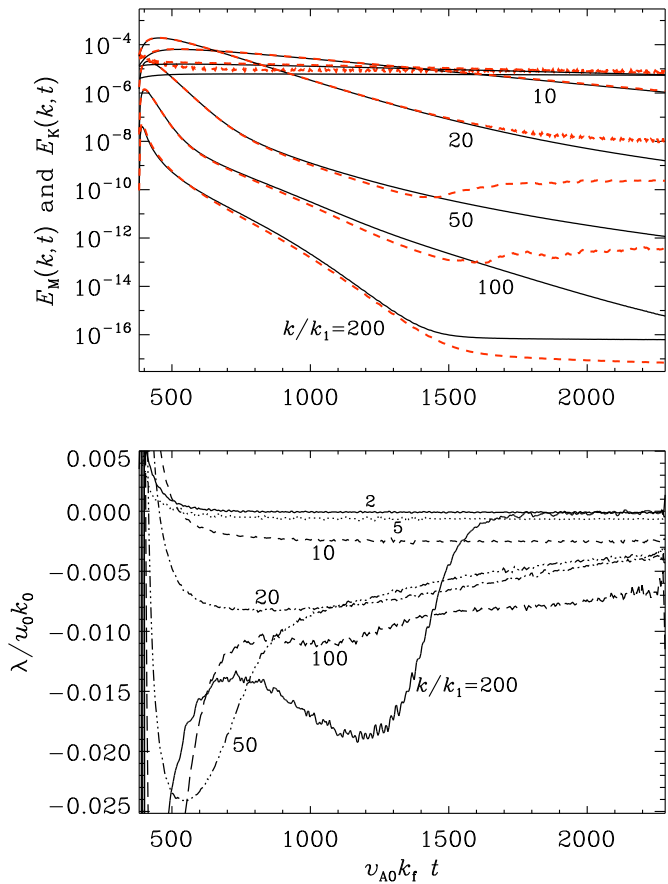}
\end{center}\caption[]{
  Upper panel: decay of magnetic energy (black lines) and kinetic
  energy (red dashed lines) at different wavenumbers
        ($ k/k_1=2$, 5, 10, 20, 50, 100, 200).
  Lower panel: instantaneous decay rates at the same wavenumbers.
}\label{pktexp}\end{figure}

\begin{figure}[t!]\begin{center}
\includegraphics[width=\columnwidth]{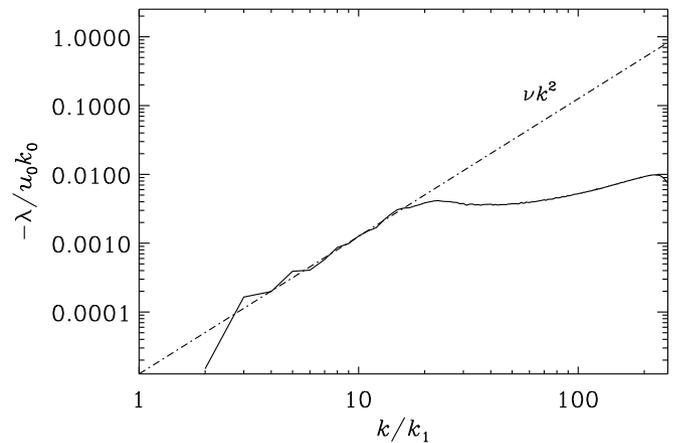}
\end{center}\caption[]{
  Instantaneous decay rates as a function of wavenumber, compared with
  the visco-resistive decay rate (dash-dotted straight line).
}\label{plamk}\end{figure}

In \Fig{pktexp} we show the decay of magnetic and kinetic energies for
a few selected wavenumbers.
The instantaneous decay rates $\lambda(t)$ are shown in the lower panel.
We determine $\lambda(t,k)$ as an average of the instantaneous decay rates
for a suitable time interval when the decay is indeed exponential.
The resulting decay rates $\overline\lambda$ are shown in
\Fig{plamk} and compared with the visco-resistive decay rate.

Note that, within a certain wavenumber interval,
the decay does indeed follow a mode-by-mode exponential
decay, so there is no coupling between different wavenumbers,
except for $k/k_1>20$, where the decay is slower than what is
expected based on a purely visco-resistive decay.
This must be a consequence of a depletion of the $\uu\times\BB$
nonlinearity when $\uu\cdot\BB$ is maximized.

        \section{Conclusions}

Our work has shown that in forced MHD turbulence with strong cross
helicity, kinetic energy displays a slight uprise at the smallest
wavenumbers.
This is reminiscent of the inverse transfer found in turbulence with
a non-Galilean invariant forcing, giving rise to an AKA effect.
Here, however, the forcing is Galilean invariant because our forcing
function is $\delta$ correlated in time, so it has no memory of the
previous forcing time step and therefore no proper motion of
the forcing field can be defined.
We are therefore tempted to associate this small uprise of spectral
kinetic energy with the presence of cross helicity.

In the decaying case, we find that the kinetic energy decays at all
wavenumbers, i.e., there is no evidence for inverse transfer or
inverse cascade behavior.
This raises doubts about our tentative conclusion regarding the forced
case, where the small uprise might also be just the result of the
finiteness of the domain and would disappear at larger scale separation
or for larger domains.
We have not studied this here, but refer instead to a similar observation
by \cite{YB16}, who simulated rotating forced hydrodynamic turbulence
in the presence of a profile of kinetic helicity and found inverse
transfer, but only when the domain was not too large.

Unlike $\EK(k,t)$, which shows a decay at all $k$, $\EM(k,t)$ does
actually show a slight increase at the smallest wavenumbers.
This is likely a consequence of the presence of finite kinetic energy
at large-scale scales
and is reminiscent of the inverse transfer found
in nonhelical MHD turbulence found in the magnetically dominated case
\citep{BKT15}, where the magnetic energy has a $k^4$ subinertial range,
which is steeper than the $k^2$ subinertial range found for kinetic
energy.
Again, then, the kinetic energy exceeds the magnetic energy at the
smallest wavenumbers.

In all the cases presented here, the relation with `ordinary' (i.e.,
low cross helicity) turbulence is unclear,
because the presence of strong cross helicity implies a suppression
of nonlinearity \citep[e.g.,][]{DobrowolnyEA80-prl}.
As a result, the decay law is, within a certain wavenumber interval,
purely exponential and not like a power law, as in ordinary turbulence.
As is well-known, when the nonlinear terms are negligible each Fourier
mode will undergo exponential decay at the appropriate
wavenumber-dependent decay rate associated with the linear dissipation
terms.

When we started our work, we regarded an initial integral wavenumber of
$15\,k_1$ as large enough.
This may need to be reconsidered in future work, because in related
studies of decaying hydromagnetic turbulence, an initial scale separation
of 60 has meanwhile been found to be more adequate \citep{BKT15}.
Also, a resolution of $512^3$ mesh points may not be enough for studying
the possibility of inverse transfer.
Most importantly, perhaps, it would be useful to work with more controlled
initial conditions that have well determined sub-inertial and inertial
range spectra.
Such studies have been done in the presence of magnetic helicity
\citep{BK17,BKMRPTV17}, but not yet in the presence of cross helicity.
Also, our way of initializing cross helicity by setting $\AAA\propto\UU$ may
be rather special.
Another possibility is to drive cross helicity through the application of
gravity and a magnetic field aligned with it, as done in the earlier work
of \cite{RKB11} and \cite{BGJKR14}.
This would lead to a stratified system in which the occurrence of inverse
transfer can directly be associated with the formation of spots that
have been found to form as a generic result of the negative effective
magnetic pressure instability for vertical magnetic fields see
Fig.~1 of \cite{BKR13} and \cite{BRK16} for a review.

\section*{\scriptsize{Acknowledgements}}
This research was supported in part by the Astronomy and Astrophysics
Grants Program of the National Science Foundation (grant 1615100),
and the University of Colorado through
its support of the George Ellery Hale visiting faculty appointment.
We acknowledge the allocation of computing resources provided by the
Swedish National Allocations Committee at the Center for Parallel
Computers at the Royal Institute of Technology in Stockholm.



\end{document}